# Ancestry-Adjusted Polygenic Risk Scores for Predicting Obesity Risk in the Indonesian Population

*Jocelyn Verna* Siswanto [1], *Belinda* Mutiara [1], *Felicia* Austin [1], *Jonathan* Susanto [1,2], *Cathelyn Theophila* Tan[1,3], *Restu Unggul* Kresnadi [1], *Kezia* Irene [1]

[1]PT Kalbe Farma Tbk, 10510 Central Jakarta, Indonesia
[2]Universitas Multimedia Nusantara, 15811 Tangerang, Indonesia
[3]Simon Fraser University, Burnaby, BC, Canada

**Abstract.** Obesity prevalence in Indonesian adults increased from 10.5% in 2007 to 23.4% in 2023. Studies showed that genetic predisposition significantly influences obesity susceptibility. To aid this, polygenic risk scores (PRS) help aggregate the effects of numerous genetic variants to assess genetic risk. However, 91% of genome-wide association studies (GWAS) involve European populations, limiting their applicability to Indonesians due to genetic diversity. This study aims to develop and validate an ancestry adjusted PRS for obesity in the Indonesian population using principal component analysis (PCA) method constructed from the 1000 Genomes Project data and our own genomic data from approximately 2,800 Indonesians. We calculate PRS for obesity using all races, then determine the first four principal components using ancestry-informative SNPs and develop a linear regression model to predict PRS based on these principal components. The raw PRS is adjusted by subtracting the predicted score to obtain an ancestry adjusted PRS for the Indonesian population. Our results indicate that the ancestry-adjusted PRS improves obesity risk prediction. Compared to the unadjusted PRS, the adjusted score improved classification performance with a 5% increase in area under the ROC curve (AUC). This approach underscores the importance of population-specific adjustments in genetic risk assessments to enable more effective personalized healthcare and targeted intervention strategies for diverse populations.

## I. INTRODUCTION

Obesity has emerged as an important worldwide health concern, with its incidence rising considerably over the last few decades. In Indonesia, the obesity epidemic poses unique public health challenges due to its rapid increase. Indonesian Baseline Health Research has highlighted the alarming rise in obesity rates among adults [1]. In 2007, 10.5% of Indonesian people aged 18 years or older were obese. By 2013, this figure had risen to 15.4%, and the most recent data from 2023 show a further increase to 23.4% [1].

While lifestyle factors such as diet, physical activity, and urbanization are well-recognized contributors to obesity, there is increasing evidence of a significant genetic component influencing individual susceptibility to obesity [2]. GWAS has identified numerous single nucleotide polymorphisms (SNPs) associated with body mass index (BMI) and obesity-related traits. For instance, variants in genes such as FTO, MC4R, and TMEM18 have been consistently linked to obesity risk in various populations [3, 4, 5]. Polygenic Risk Scores (PRS) can be calculated using these SNPs to estimate an individual's genetic predisposition to obesity [6].

However, around 91% of these studies have been conducted in populations of European descent, raising concerns about the relevance and applicability of these findings to non-European populations [7]. The allele frequencies and effect sizes of obesity-associated SNPs may differ across populations due to genetic diversity and environmental interactions [8]. Consequently, PRS developed from European-centric GWAS may not accurately predict obesity risk in non-European populations, including those in Indonesia [9, 10, 11]. This underlines the importance of PRS ancestry adjustment in diverse populations to ensure that genetic risk predictions are accurate and applicable across different ethnic groups. To address these issues, we utilized Principal Components Analysis (PCA) to improve the applicability of genetic risk predictions for obesity in Indonesian populations. PCA is a statistical method that transforms high-dimensional genetic data into a smaller set of uncorrelated variables, called principal components, which capture the major axes of genetic variation within a population [12]. PCA-based ancestry adjustment has been widely adopted in genetic epidemiology to enhance the generalizability of risk predictions across ethnic groups [13].

## II. RELATED WORKS

A study by Ducan et. al [14] conducted a comprehensive analysis of PRS performance across various ancestries. They found that scores derived from European data exhibit significantly lower predictive power in non-European groups, such as African and Asian populations (PRS Performance). This reduced accuracy is attributed to variations in linkage disequilibrium and allele frequencies, which differ markedly across populations. Similarly, Martin et al. [11] highlighted that the clinical application of unadjusted PRS could exacerbate health disparities, particularly in underrepresented populations, underscoring the need for population-specific genetic risk assessments (Health Disparities).

A study by Hae-Un Jung [15] highlighted that even with cross-population PRS approaches, East Asian PRSs still lag behind with their European counterparts in predictive performance for certain traits. The authors emphasized that direct application of European PRSs to East Asian Populations remains challenging due to differences in linkage disequilibrium (LD) patterns, allele frequencies, and genetic architectures.

A study by Peh Joo Hoo et al. [16] evaluated the utility of PRS for common cancers in large East Asians cohort and found that while PRSs can effectively stratify cancer risk, their predictive performance remains suboptimal when applied to non-European populations. The authors emphasized that this reduced accuracy is largely due to the fact that most PRSs were developed using predominantly European GWAS data, which fails to capture the genetic diversity present in Asian populations. The authors argue that improving PRS performance for non-European populations will require more diverse training datasets that better account for the genetic architecture of these populations, reinforcing the critical need for ancestry-adjusted PRS.

Despite advances in PRS for obesity, significant gaps remain in their application to non-European populations, particularly Indonesians. While studies in other Asian populations, like Koreans, have shown promise in developing population-specific PRS (Korean PRS), genetic research on obesity in Indonesia is scarce. The lack of Indonesian-specific genomic data and ancestry-adjusted PRS exacerbates health disparities, as unadjusted scores may misclassify risk in this population. The current study addresses these gaps by developing and validating an ancestry-adjusted PRS using data from the 1000 Genomes Project and approximately 5800 Indonesian individuals to improve obesity risk prediction and support personalized healthcare in an underrepresented population.

## III. METHODS

### A. Dataset

In this research, we utilized the 1000 Genomes Project data as a training dataset and our in-house targeted genome sequencing Indonesian data as the testing dataset. Both datasets consist of genomic data from more than 7000 individuals across various populations worldwide and include both male and female samples. More detailed population distribution can be seen in Table 1. This diverse dataset provides a comprehensive reference for human genetic variation.

**Table 1.** Data Demography

| Characteristic | 1000 Genome Data | Indonesian Data |
|---|---|---|
| Sample Size (N) | 2504 | 4628 |
| Genotyping Platform | Illumina [17] | Kalgen01 Chip on GeneTitan |
| Sampling Year | 2008 - 2010 | 2021 - 2024 |
| Sex (n, %) | | |
| Male | 1740 (49.07%) | 1007 (21.76%) |
| Female | 1761 (50.30%) | 2159 (46.65%) |
| Unknown | N/A | 1462 (31.59%) |
| Ethnic Breakdown (n, %) | | |
| European | 503 (20.09%) | N/A (Needs Adjustment) |
| African | 661 (26.40%) | |
| East Asian | 504 (20.13%) | |
| South Asian | 489 (19.53%) | |
| Hispanic | 347 (13.86%) | |
| Age at Recruitment (years) | | |
| Mean (Std) | N/A (All above 18) | 38.52 (10.52) |
| Min – Max | | 18 – 65 |
| Median | | 37 |

### B. Methods

We utilized methods from Khera et al. [18] and Hao et al. [19] to develop our ancestry adjusted PRS algorithm. Figure 1 shows our comprehensive workflow.

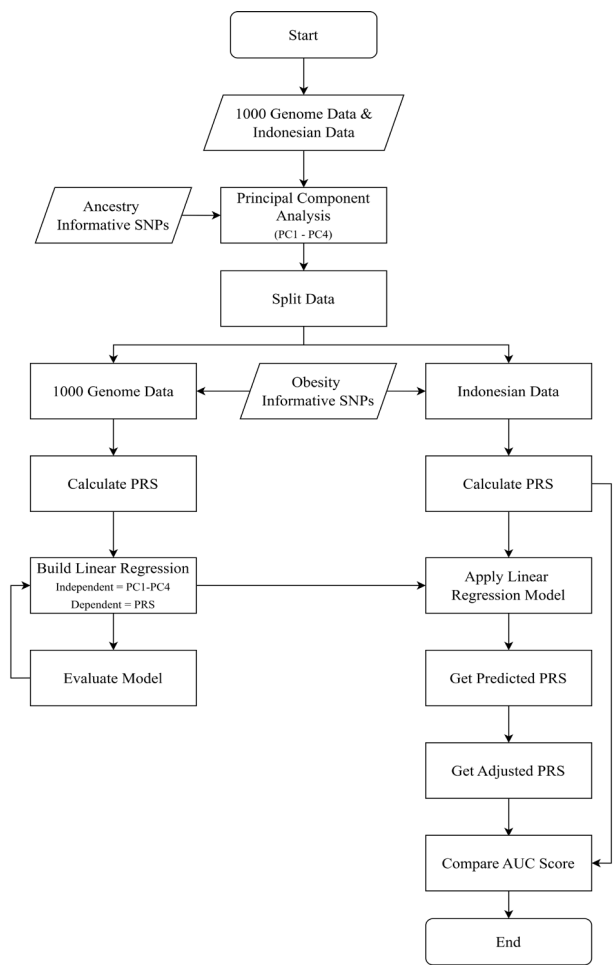

**Figure 1** Our Method Diagram

We divided the data into training and testing datasets, where the 1000 Genomes Project data was used for training and Indonesian data for testing. On each dataset, we filtered the SNPs using a set of 16,385 previously reported ancestry-informative SNPs [19]. Then, we filtered the obesity-related SNPs based on the research of Yoon and Cho [9].

We built the PCA based on the ancestry-informative SNPs data to capture genetic ancestry information while reducing data dimensionality and calculated the raw PRS using the obesity SNPs data. PCA creates a new set of variables by converting the initial variables into principal components (PCs), which are linear combinations of the original variables. Mathematically, this involves computing the eigenvectors and eigenvalues of the covariance matrix of the standardized genotype data. Let $X$ represent the standardized genotype matrix (with individuals as rows and SNPs as columns). The covariance matrix $C$ is computed as:

$$C = \frac{1}{n-1} X^T X \quad (1)$$

where $n$ represents the quantity of samples. The following eigenvalue problem is then resolved by PCA:

$$Cv = \lambda v \quad (2)$$

where $\lambda$ is the associated eigenvalue, which is known as the variance explained by that component, and $v$ an eigenvector, or the direction of a primary component. The transformed data can be obtained as:

$$Z = XW \quad (3)$$

where $W$ is the matrix of selected eigenvectors and $Z$ is the lower-dimensional representation of the initial data. [20].

After calculating PCA and raw PRS, we merged the principal components (PCs) and PRS variables based on participant's ID. Then, we built a linear regression model between the calculated PRS and the first four PCs obtained from the PCA, which cumulatively accounted for a significant portion of the variance. The model provided the projected PRS, which we subtracted from the raw PRS to get the adjusted PRS. This calculation is represented by the formula below:

$$PRS_{adj} = PRS_{raw} - (\beta_0 + \beta_1 PC_1 + \cdots + \beta_4 PC_4) \quad (4)$$

where $PRS_{adj}$ is the adjusted PRS removing the population structure effects captured by the first four principal components (PCs). $PRS_{raw}$ is the initial PRS calculated directly from the obesity-related SNPs without accounting for population structure. $\beta_0$ (Intercept) represents the intercept term of the linear regression model that captures the baseline effect of the population structure on the raw PRS, independent of the principal components. The $\beta_1$ - $\beta_4$ (PC's Coefficients) represent the effect of each PC on the raw PRS, these are estimated during the training phase of the linear regression model. $PC_1 - PC_4$ represent the individual-specific values (scores) on the first four principal components, which capture the major axes of genetic variation. These values are derived from PCA conducted on ancestry-informative SNPs and reflect each individual's genetic ancestry profile.

The same analysis was then applied to our testing dataset. We first gathered the Indonesian patient data, filtered the ancestry-informative SNPs, then applied PCA out of it. Before the SNPs filtering is done, we performed genotype imputation to address the limited coverage of ancestry-informative SNPs. Our dataset initially contained a fraction of the required 16,385 ancestry-informative SNPs. Thus, we used Beagle, an imputation tool developed by Brian L. Browning and Sharon R. Browning, which employs a haplotype-based Hidden Markov Model (HMM) [21]. Beagle infers missing genotypes by matching observed haplotypes to a reference panel, in this case from the 1000 Genomes Project, to statistically reconstruct likely missing variants. Additional information for our imputation method will be published in our subsequent manuscript.

Once imputation is done, we extracted the ancestry-informative SNPs and performed PCA to capture population structure. We also filtered the obesity-related SNPs specific to the Indonesian population to calculate the raw PRS for obesity. Using the previously constructed linear regression model, we predicted the PRS for each Indonesian individual based on their first four PCs scores. We then obtained the adjusted PRS by subtracting the predicted PRS from the raw PRS.

After retrieving the adjusted PRS, we evaluate its predictive utility by constructing a linear regression model where the adjusted PRS served as the independent variable and obesity status (obese vs. non-obese) was treated as the dependent variable. The obesity classification was based on ground truth labels derived from our Indonesian cohort, where the ground truth labels for obesity were derived by classifying individuals based on a body mass index (BMI) threshold of greater than 27, which aligns with obesity criteria of Indonesian populations [1]. Individuals with a BMI below this threshold were labelled as non-obese.

In parallel, we developed a comparison model using the raw (unadjusted) PRS as the independent variable. By comparing the performance of these two models based on area under the receiver operating characteristic curve (AUC), we aimed to determine the extent to which adjusting the PRS improved the model's ability to predict obesity status.

## IV. RESULTS & DISCUSSION

We first built PCA using 16,385 SNPs and determined the number of Principal Components (PCs) necessary to represent the variance from the ancestry-informative SNPs. Using the explained variance ratio curve, we found that the first four PCs able to explain more than 80% of total variance.

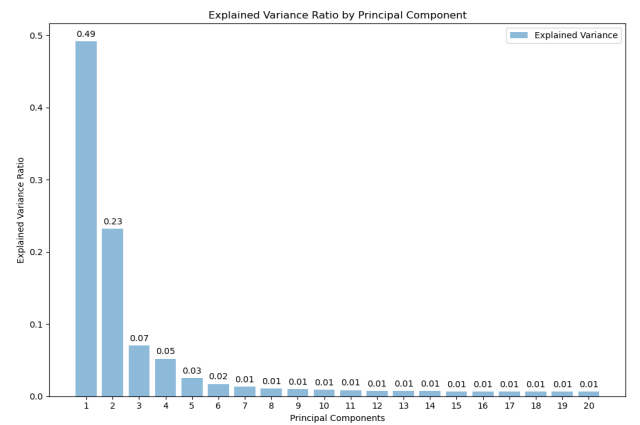

**Figure 2** Explained Variance Ratio of PC1-PC20

However, after the 4th PC, the ratios drop below 5% and become flat, suggesting no significant improvement in explained variance. The cumulative variance explained by the first four PCs is 83.34%, which we considered sufficient to represent the overall structure, and therefore used PC1–PC4 in subsequent analyses.

The scatter plot on Figure 3 and Figure 4 visualizes ethnicity distribution of all participants, where each ancestry is shown in different colors. Figure 3 uses PC1 vs. PC3 to plot the distribution, whereas Figure 4 uses PC2 vs. PC3. Overall, we can see on the plot that the individuals of the same ancestry (e.g., East Asian) are grouped together, demonstrating that the PCA effectively reduces dimensionality while also retaining essential information. Moreover, we can also see that Indonesian people appear to form a separate but proximal cluster near the Asian groups, particularly East Asian. This suggests there may be shared genetic background or intermediate ancestry patterns.

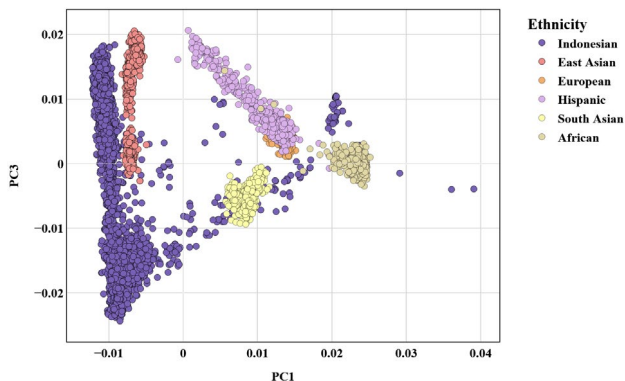

**Figure 3** Result of PC1 vs PC3 by Ethnicity

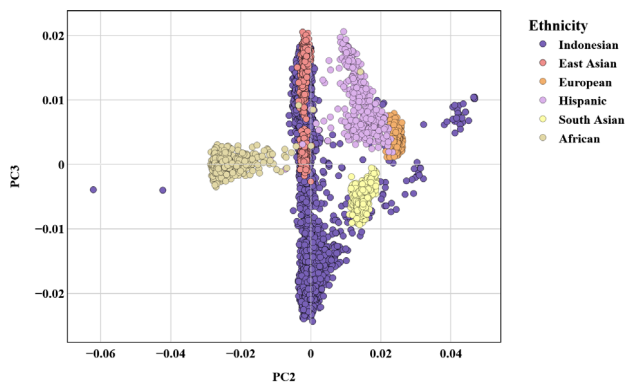

**Figure 4** Result of PC2 vs PC3 by Ethnicity

After calculating both raw and adjusted PRS based on obesity-associated SNPs, we examined the distribution of PRS values across different populations.

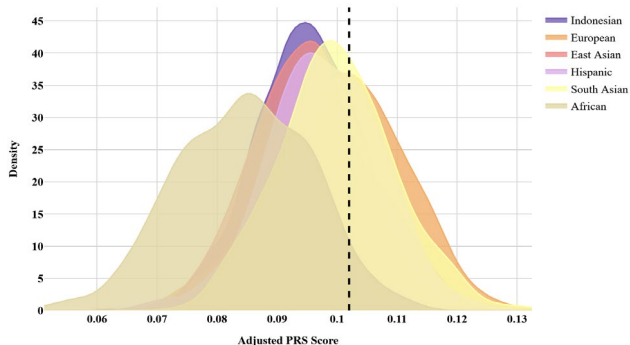

**Figure 5** PRS Distribution Before Adjustment

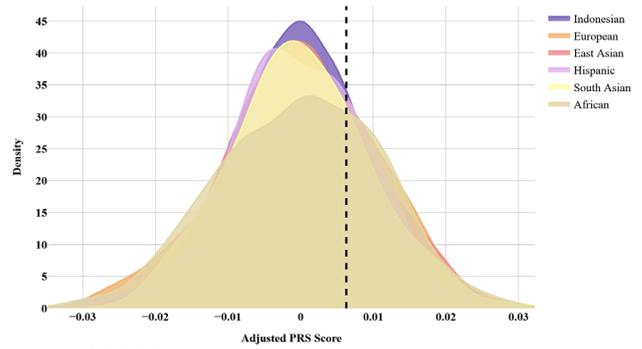

**Figure 6** PRS Distribution After Adjustment

In Figure 5, the PRS values vary quite significantly between populations, which may reflect biases in the genetic data. After adjustment, the distributions become more similar between populations, as seen in Figure 6. This indicates that the adjustment helps to normalize the PRS value, reducing population-specific biases. By improving consistency in PRS interpretation, the adjustment process helps mitigate the risk of overestimating or underestimating obesity risk in certain ancestries and supports more reliable use of PRS in diverse populations.

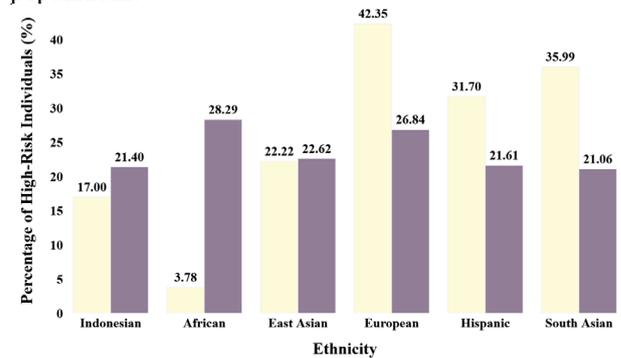

**Figure 7** Proportion of Obesity-Risked Individual by Ethnicity

Figure 7 presents a stacked bar chart showing the proportion of individuals classified as high risk for obesity based on the 76th percentile of both raw and adjusted Polygenic Risk Scores (PRS) across multiple ethnic groups. Based on the chart result, using raw PRS tends to overestimate the proportion of high-risk individuals in Europeans and South Asians cohort, while underestimating risk in Africans and Indonesians.

After applying ethnic-specific adjustments to the PRS, the distribution of high-risk individuals becomes more balanced and reflective of actual population-level risk profiles. For instance, the adjusted PRS in Indonesian cohort identified 21.4% of individuals as high-risk, which closely aligned with national data from Riskesdas [1] that reports a recent obesity prevalence of 23.4% in Indonesia. This concordance lends support to the validity of the adjustment approach and underscores the importance of adjusting PRS to account for ancestral diversity.

To further assess the predictive performance of the PRS, we evaluated the ability of both raw and adjusted PRS to classify individuals as obese or non-obese using receiver operating characteristic (ROC) analysis. Based on Figure 8, the AUC for the raw PRS was 0.57, which was outperformed by the adjusted PRS that generated AUC of 0.57. While both models perform modestly above random chance (AUC = 0.5), an improvement in the adjusted PRS demonstrates the model's enhanced discriminatory power.

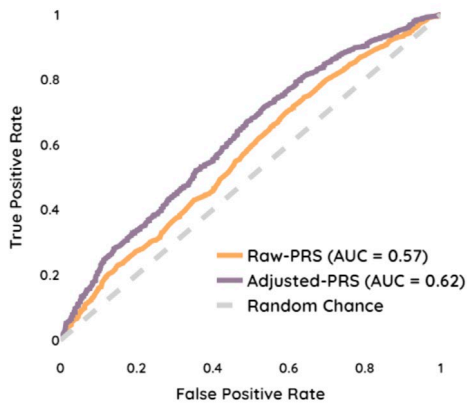

**Figure 8** AUC of Raw PRS vs Adjusted PRS

The increase in AUC highlights the value of adjusting PRS to better account for population-specific characteristics and potential confounders. The adjusted model not only aligns better with observed obesity prevalence in Indonesia [1] but also improves the model's ability to correctly identify individuals at high risk. These results support the importance of PRS refinement in improving the clinical relevance and accuracy of genetic risk prediction in diverse populations.

## V. CONCLUSION

In this study, we demonstrated the importance of adjusting Polygenic Risk Scores (PRS) to account for population-specific genetic diversity. By comparing the distribution and predictive model's AUC score of raw PRS and ancestry adjusted PRS, we showed significant improvements in the accuracy of obesity risk predictions for the Indonesian population. The use of ancestry adjusted PRS enhances the precision of personalized healthcare and intervention strategies, facilitating more effective management and prevention of obesity. Without such adjustments, genetic risk predictions may lead to misclassification and potential health disparities, especially in underrepresented populations.

Given that these results are still preliminary, the future plans may include replication with larger sample sizes to enhance the robustness of our findings. Other than that, the obesity ground truth used for classification was based solely on BMI thresholds, which may lack clinical precision and fail to capture the complexity of obesity as a condition. Clinical and functional validations may be essential to confirm the ground truth and assess the practical applicability of our adjusted PRS accurately. Additionally, our model focused exclusively on genetic predictors and did not account for environmental or lifestyle factors such as diet, physical activity, and socioeconomic status, which are commonly known to interact with genetic predisposition. Future studies can involve integrating additional layers of data, such as environmental and lifestyle factors, to develop more comprehensive models for obesity risk assessment.

This approach not only benefits the Indonesian population but also sets a precedent for future research to develop and validate ancestry adjusted PRS across diverse populations. Such efforts are crucial in addressing global health disparities and ensuring that genetic risk assessments are inclusive and applicable to all ethnic groups. Our findings underscore the need for more research focusing on minority populations, such as Indonesians and other Southeast Asian people, in order to improve global health outcomes and advance personalized medicine.